\documentstyle[prb,aps,amsfonts,amssymb,epsfig,floats]{revtex}
\begin{document}
\draft
\preprint{}
%
%following two lines are for two column format
\twocolumn[\hsize\textwidth\columnwidth\hsize\csname
@twocolumnfalse\endcsname
\title{Vortex melting and decoupling transitions in YBa$_{2}$Cu$_{4}$O$_{8}$ single
crystals}
\author{X.G. Qiu}
\address{Laboratorium voor Vaste-Stoffysica en Magnetisme, Katholieke Universiteit
Leuven, Celestijnenlaan 200D, B-3001 Leuven, Belgium and \\
Photodynamics Research Center, the Institute of Physics and Chemistry
(RIKEN), 19-1399 Koeji, Nagamachi, Aoba-ku, Sendai 980-0868, Japan }
\author{V.V. Moshchalkov}
\address{Laboratorium voor Vaste-Stoffysica en Magnetisme, Katholieke Universiteit
Leuven, Celestijnenlaan 200D, B-3001 Leuven, Belgium}
\author{J. Karpinski}
\address{Laboratorium fur Festkorperphysik, ETH-Honggerberg, CH-8093, Zurich,
Switzerland}

\date{\today}

\maketitle

\begin{abstract}
The vortex correlation along the c-axis in high quality single crystals of
YBa$_{2}$Cu$_{4}$O$_{8}$ has been investigated as a function of temperature
T in different magnetic fields, using the quasi-flux transformer
configuration. A simultaneous sharp drop associated with the vortex lattice
melting is observed in both the primary and secondary voltages(V$_{top}$ and
V$_{bot}$). Just above the melting temperature, the vortices form
three-dimensional line liquid with the correlation length along the c
direction $L_{c}\leq $ t, the sample thickness. The temperature where a
resistive peak in R$_{bot}$ develops corresponds to the decoupling
temperature T$_{d}$ at which the vortices loose their correlation along the
c-direction and they dissolve into the two dimensional pancake vortices. The
H-T phase diagram for the YBa$_{2}$Cu$_{4}$O$_{8}$ single crystal is
obtained.
\end{abstract}
\vskip2pc]

\narrowtext

\section{Introduction}

The vortex dynamics in the mixed state of high T$_{c}$ superconductors
(HTSCs) remains a topic subject to intense investigation because of its
importance for both the fundamental research and future applications. It is
now well established that the thermal fluctuation induces a second order
transition from a high temperature vortex liquid phase to a low temperature
vortex glass phase for the vortex matter in a superconductor with a strong
disorder. In a clean superconductor the second order transition is replaced
by a first order transition from a vortex liquid to an Abrikosov vortex
lattice.\cite{ref1}

Flux melting in various high quality high T$_{c}$ superconducting single
crystals has been observed by different experimental techniques.\cite
{ref2,ref3,ref4,ref5,ref6,ref7,ref8} It demonstrates itself as a sharp
resistive drop or a jump in the magnetization. An important issue concerning
the melting transition is that if the vortices lose their coherence in the c
direction during the melting transition or they will keep the c-axis
correlations intact and then lose them at a still higher temperature.\cite
{ref9} This issue has been widely pursued by doing transport measurements
with the dc flux transformer configuration. So far the results remain
controversial. Doyle et al.\cite{ref10} and Fuchs et al.\cite{ref11} found a
coincidence of the melting transition and a decoupling one on Bi$_{2}$Sr$%
_{2} $CaCu$_{2}$O$_{8}$ (Bi2212) single crystals. However, Wan et al.\cite
{ref12} and Keener et al. \cite{ref13} did similar measurements on Bi2212
single crystals. They concluded that the melting transition took place in a
two-stage fashion, i.e., the rigid lattice at low temperature first melts
into a three dimensional (3D) liquid followed by a decoupling transition
upon an increase in the temperature. Recently Blasius et al.\cite{ref14}
performed $\mu $-spin rotation measurements on Bi2212 single crystals with
different oxygen contents and they obtained evidence for a two stage
transition of the vortex matter as a function of temperature under
equilibrium conditions. On the other hand, in the YBa$_{2}$Cu$_{3}$O$%
_{7-\delta }$ (Y123) system a coincidence of vortex melting and loss of
vortex correlation along the c-axis was observed by L\'{o}pez et al.\cite
{ref15} A decoupling transition was not observed in the optimally doped Y123
system since the vortex in this system is always in the 3D region.

Underdoped YBa$_{2}$Cu$_{4}$O$_{8}$ (Y124) is naturally stoichiometric and
untwinned, it has a moderate anisotropy parameter $\gamma $ as compared to
those of Y123 and Bi2212. The weakness of the pinning strength in this
material reduces the contribution from spurious effects in the melting
transition in the vortex system. Thus it is a good prototype material for
the investigation of the interplay between the vortex interaction and
dimensionality on the vortex dynamics. Recently, we observed a first order
melting transition in the Y124 single crystals. \cite{ref16} It is of great
interest to extend our previous work to examine the vortex dynamics along
the c-axis. Here we report the transport data obtained in simultaneous
measurements for both the primary (V$_{top}$) and secondary (V$_{bot}$)
voltages using the dc transformer configuration, in order to investigate the
vortex correlation along the c-axis in the Y124 system. It is demonstrated
that in Y124 single crystals, vortex lattice (VL) melts into a 3D vortex
liquid via a first order transitionand and that the decoupling transition is
seperated from the melting one. Our results support the two stage transition
scenario that the vortices first undergo a melting transition, followed by a
disappearance of correlation in the transverse direction at a higher
temperature.

\section{Experimental and Results}

YBa$_{2}$Cu$_{4}$O$_{8}$ single crystals were grown by the high-pressure
flux method as described previously.\cite{ref17} These single crystals were
needle-like with typical dimensions of 1.2$\times $0.4$\times $0.01 mm$^{3}$%
. The zero field critical transition temperature T$_{c}$(0) of the single
crystals was about 78 K. Two single crystals were used for these
measurements. Each crystal was carefully cleaved to obtain optically flat
surfaces with the c-axis normal to the sample surface. Gold wires were
attached to the top and bottom surfaces of crystal by using Platinum epoxy.
Then the crystal was heated in air at 100 ${^{\circ }}$C for 1 hour, the
resultant contact resistance was typically below 0.5 $\Omega $. The
electrical contact geometry for the measurement is shown in the inset of
Fig. 1. The secondary voltage contact pair (5 and 6) on the bottom ab face
was placed directly beneath the top face primary voltage pair (2 and 3). The
distance between each voltage pair was about 0.5 mm. The resistance were
measured using a low frequency (17 Hz) ac lock-in technique with an
excitation current of 0.1 mA injected from contact 1 to 4 while measuring
simultaneously the voltage drops V$_{23}$ and V$_{56}$ with the help of a
Keithley 228 scanner. The magnetic field was generated by a 15 Tesla Oxford
superconducting magnet and was applied parallel to the c-axis of the crystal
throughout the measurements.

\begin{figure}[bht]
\centerline{
\epsfxsize=3.2 in
\epsfbox{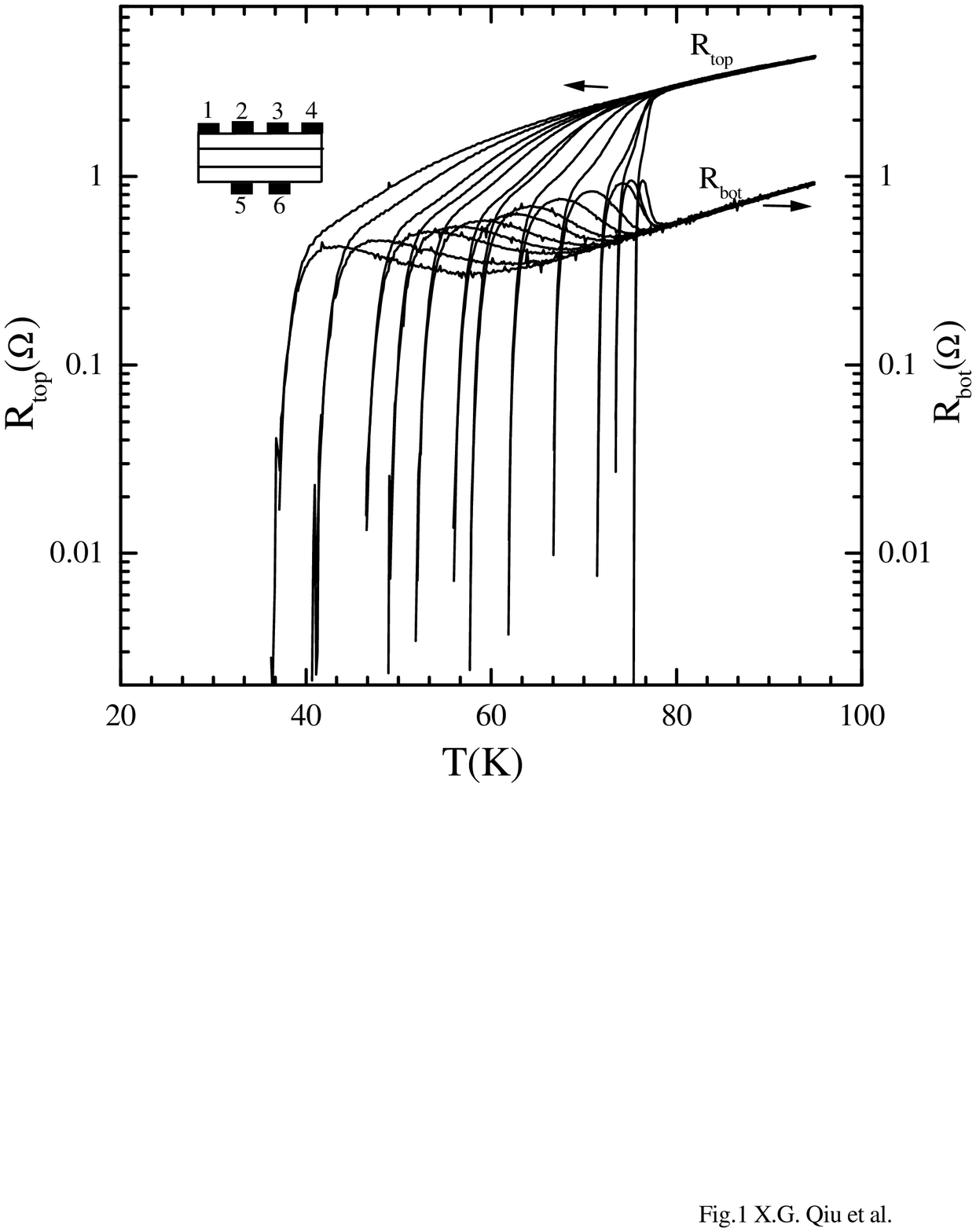}}
\caption{The logarithmic plot of the temperature dependence of R$_{ab}$ and R%
$_{c}$ at different applied magnetic fields. From right to left: H=0.2, 0.5,
1, 2, 3,4, 6,7, 8, 10, 12 T. Inset: the electrical contact configuration for
the quasi flux transformer measurements.}
\label{fig1}
\end{figure}

Shown in Fig. 1 are the representative R$_{top}$(T) and R$_{bot}$(T)
(defined as V$_{23}$/I$_{14}$ and V$_{56}$/I$_{14}$, respectively) curves
taken at different applied magnetic fields up to 12 Telsa for one of the
Y124 single crystals. Similar results were obtained for the other crystal.
The resistivity was measured by first cooling down the sample below T$_{c}$
and then collecting the data while warming up the sample. While the applied
magnetic fields induce a negligible depression in the superconducting onset
temperature for the in plane superconductivity, they appear to considerably
suppress the temperature where R$_{bot}$ deviates from its normal state
value. A sharp jump in R$_{top}$ with a magnitude of R/R$_{n}$ $\sim $ 10\%
(R$_{n}$ being the normal state resistance) is clearly observed at each
magnetic field. We attribute it to the occurrence of the vortex melting
transition. We define the melting transition temperature T$_{m}$ as the one
where a sharp peak in dR/dT exists. The temperature T$_{m}$ defined by R$%
_{top}$(T) and R$_{bot}$(T) are equal, i.e., T$_{m}^{top}$=T$_{m}^{bot}$.
This is demonstrated more clearly by plotting the derivatives of R$_{top}$
and R$_{bot}$ with respect to T together, as shown in Fig. 2a. Below the
melting temperature, R$_{top}$=R$_{bot}$. Immediately above the melting
temperature T$_{m}$, R$_{top}$ becomes larger than R$_{bot}$. As the
temperature increases, both R$_{top}$ and R$_{bot}$ increase and the
difference between R$_{top}$ and R$_{bot}$ becomes larger and larger.
Finally near a characteristic temperature T$_{d}$, R$_{bot}$ reaches its
maximum and at the same temperature, a small kink develops in R$_{top}$, as
can be seen in a separated plot in Fig. 2b. It is interesting to notice that
above T$_{m}$, dR$_{top}$/dT decreases with the increasing temperature,
reaches a minimum at T$_{d}$ before a second local maximum appears at T$_{p}$%
. At exactly the same temperature T$_{p}$, dR$_{bot}$/dT shows a local
minimal. This behavior persists for the dR/dT curves at all applied magnetic
fields. By plotting T$_{m}$ and T$_{d}$ as a function of the magnetic field,
we obtain a phase diagram for this Y124 single crystal as shown in Fig. 3.
Previously we have found that T$_{m}$(B) could be well described by the
anisotropic Ginzburg-Landau theory and the melting line could be fitted by
an empirical formula B$_{m}$(T)=31.4(1-T/T$_{c}$)$^{1.44}$ with T$_{c}$=78.6
K.\cite{ref16} The fitting result is shown as the solid line in Fig.3. In
the mean time, angular dependence of the melting transition gives for
Lindermann criterium {\it c}$_{L}$=0.14 and the anisotropy parameter $\gamma 
$=12.4.\cite{ref16}

\begin{figure}[t]
\centerline{
\epsfxsize=3.2 in
\epsfbox{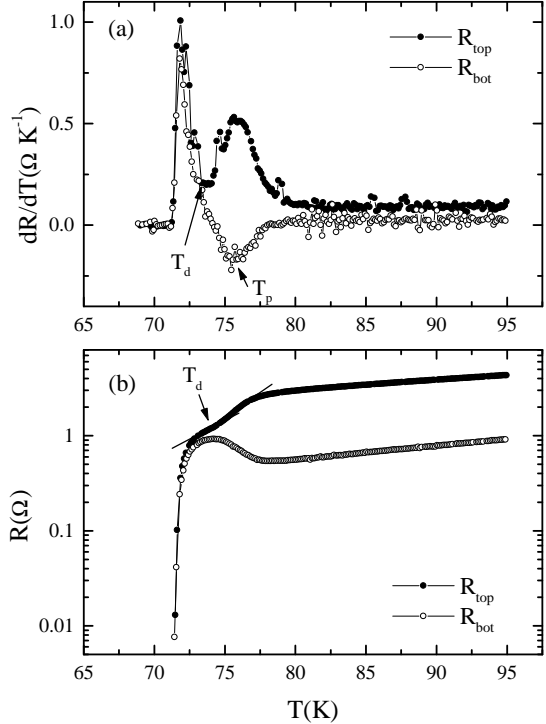}}
\caption{(a) dR$_{top}$(T)/dT vs. T and dR$_{bot}$(T)/dT vs. T at an applied
magnetic field of 1 T; (b) R$_{top}$ and R$_{bot}$ at 1 T. The temperature
at the cross point of the two solid lines is the crossover temperature
between two TAFF regions.}
\label{fig2}
\end{figure}

\begin{figure}[t]
\centerline{
\epsfxsize=3.2 in
\epsfbox{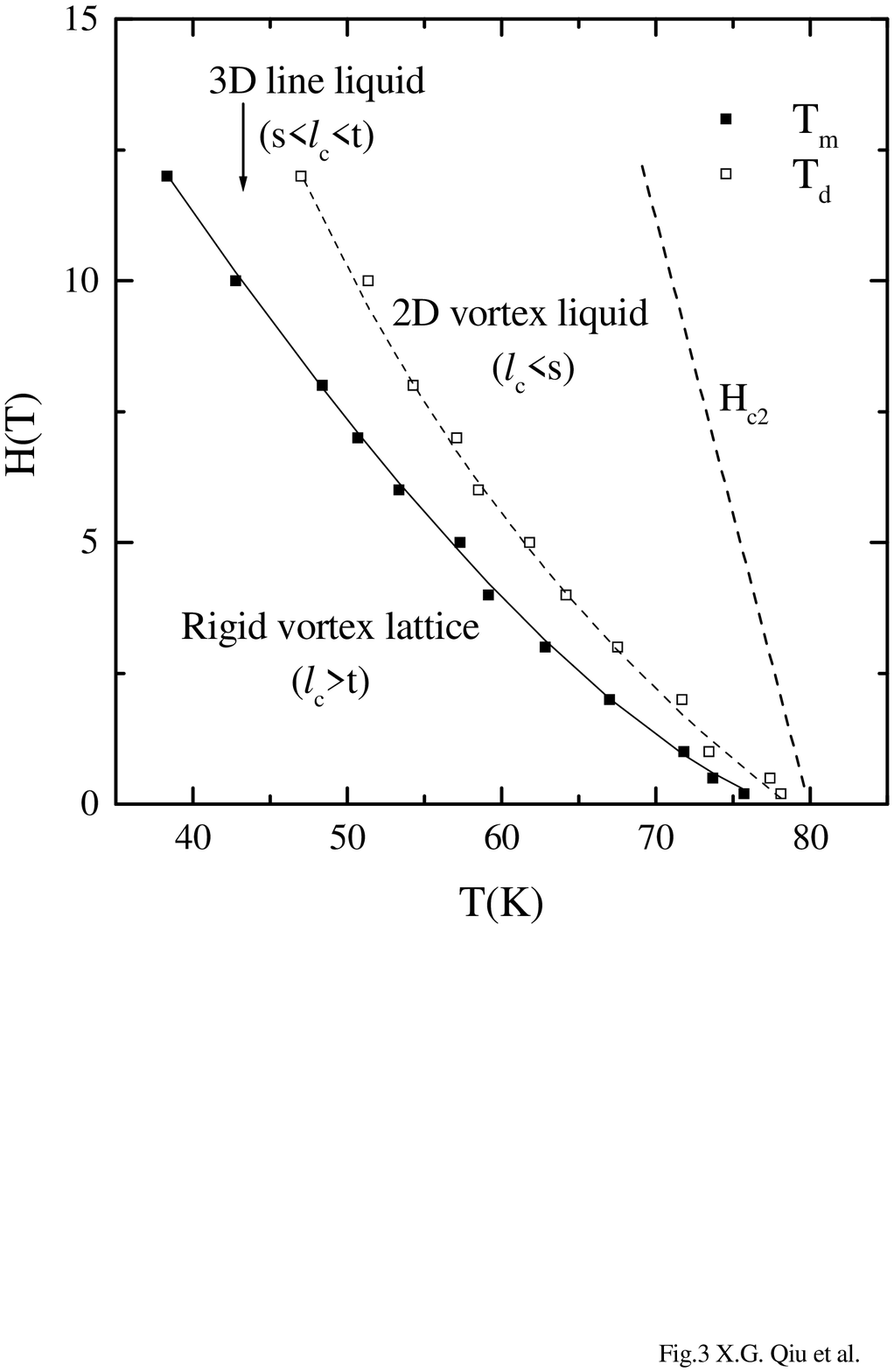}}
\caption{H-T phase diagram extracted from the transport measurements. Solid
circles are the points for the melting line, open circles are those for the
decoupling line. Solid line is the fitting result with the empirical
relation B$_{m}$(T)=31.4(1-T/78.6)$^{1.44}$. Dashed line is the decoupling
line fitted with B$_{d}$(T)=1405.4(1-T/78.6)/T. t is sample thickness and s
the distance between the neighboring CuO planes.}
\label{fig3}
\end{figure}

\section{Discussions}

The dc flux transformer configuration is useful means for probing the
dimensionality and longitudinal correlation of vortices in the mixed state.
It measures the velocity correlation for vortices running along the CuO
planes of Y124 since the voltage drops V$_{23}$ and V$_{56}$ (and thus R$%
_{top}$ and R$_{bot}$) in the mixed state is induced by the motion of
vortices whose velocity is determined by the integrated Lorentz force on the
vortex segments of length l$_{c}$ piercing the top and bottom surfaces,
respectively. The relation between R$_{top}$ and R$_{bot}$ is determined by
the velocity of the vortices in the top and bottom surfaces. When the vortex
correlation length along the c-axis is longer than the sample thickness, the
voltage drops at the top and bottom surfaces would be of the same magnitude.
The fact that immediately above the melting temperature R$_{top}$ is larger
than R$_{bot}$ suggests that the l$_{c}$ is smaller than the sample
thickness {\it t} above the melting transition.

Above T$_{c}$ in the normal state, R$_{top}$ is larger than R$_{bot}$
because the sample geometrical aspect, electrode arrangement and anisotropy
result in a nonuniform current distribution, i.e., more current will pass
through the top surface than on the bottom surface. The fact that in the
normal state the ratio R$_{top}$/R$_{bot}$ is nearly constant suggests that
the anisotropy of Y124 is nearly temperature independent in this temperature
region. Upon entering the superconducting state, the superconductivity will
make an easy path for electrons to run across the CuO planes and therefore
reduces the anisotropy. As we will discuss below, the vortices at this
temperature region is the 2D pancakes, the voltage drop will be mainly
determined by the anisotropy. Since we are measuring the resistance by
passing a constant current, the decrease in anisotropy means that more
current will pass through the CuO layers and that the current flowing in the
bottom layers will increase. The decreasing anisotropy with decreasing
temperature is reflecting in the reduction of dR$_{bot}$/dT as shown in
Fig.2a.

In the present measurement configuration, R$_{top}$ is equivalent to the
in-plane resistance R$_{ab}$ commonly measured in a four terminal
measurement. Our previous work on the in-plane resistivity demonstrated that
for R$_{top}$(T), above the melting temperature in the liquid state, there
are two distinct parts where lnR$_{top}$ shows linear dependence upon 1/T
with different slopes, which is a typical characteristic for thermally
activated flux flow (TAFF) behavior. \cite{ref16} We have identified the
temperature T$_{d}$ where the crossover between the two TAFF regions occurs
as the decoupling transition through an activation energy analysis. The
decoupling temperature is described by\cite{ref9}

\begin{equation}
B_{cr}=\frac{\Phi _{0}^{3}}{16\pi ^{3}k_{B}Tse\lambda _{ab}^{2}\gamma ^{2}}%
\text{,}  \label{eq2}
\end{equation}
where s is the interlayer spacing between the neighboring CuO planes. With a 
$\gamma $ value of 12.4 a satisfied fit is obtained by Eq. (1) for the
decoupling line shown as the dashed line in Fig. 3.

We notice that the temperature where R$_{bot}$ shows a maximum corresponds
to exactly the same temperature where a crossover between two different TAFF
regions occurs as we have identified before. Such a peak resembles that of
the out-of-plane resistance R$_{c}$ often appears in single crystals with
large anisotropy such as Bi2212.\cite{ref18,ref19} However it can be seldom
seen in single crystal with moderate anisotropy such as fully oxidized Y123.
Although there is no consensus on the occurrence of the peak in R$_{bot}$
and R$_{c}$ yet, one possible reason for it could be the tunneling of
vortices over the c-axis barrier in the liquid state following by the
complete coupling of CuO planes. Recently, there have been considerable
efforts devoted to the study of interlayer charge dynamics by far infrared
reflectivity.\cite{ref20,ref21} Studies on the interlayer charge dynamics
with the electric field component {\bf E} polarized along the c-axis reveals
that above a characteristic temperature T$_{pl}$, the superconductor behaves
like a poor metal or an ionic insulator. The carriers are confined inside
the CuO planes and the c-axis resistivity is controlled by the interlayer
tunneling. The c-axis coherence is signaled by the appearance of a
reflectivity edge associated with the Josephson plasma from the intrinsic
Josephson junctions.\cite{ref22,ref23} Above T$_{pl}$, the electron
transport along the c-axis is incoherent. Below T$_{pl}$, coherence is built
up for the interlayer carrier transport. Therefore, given the high
transition temperature and weak interlayer magnetic coupling between the
pancake vortices in the neighboring CuO planes, T$_{d}$ would quite possibly
correspond to T$_{pl}$ (the Josephson plasma frequency) at which the phase
coherence and thus the superconductivity is built up for the entire system
along the c-axis. The suppression of the onset temperature for c-axis
superconductivity by an applied magnetic field is due to the breakdown of
the phase coherence along the c-axis by a magnetic field. This is consistent
with the model proposed by Brice\~{n}o et al.\cite{ref19} and later
developed by Suzuki et al.\cite{ref24} who suggested that phase fluctuation
induced dissipation controlled the dissipation and could be well described
by the Ambegaokar-Baratoff theory.

Although the picture of vortex melting has been widely accepted, the
mechanism of the vortex melting has yet not well understood. Recently,
Nonomura et al.\cite{ref25} carried out Monte Carlo simulations of the three
dimensional frustrated XY model, in which the melting temperature was
determined by the helicity modulus along the c-axis. They found that the
melting transition is propagated through vortex entanglement. At the melting
temperature, the percentage of entangled vortices abruptly changes. Upon
increasing the temperature, the entanglement length becomes smaller than the
sample thickness and decreases rapidly. The dissipation is governed by the
cutting of vortices above the melting temperature. As a consequence of the
entanglement mechanism of the FLL melting, T$_{m}$ is scaled by the inverse
of the system size along the c axis. Our results can be explained by their
simulation results very well since we indeed observed a melting transition
into 3D line liquid with its correlation length l$_{c}$ smaller than the
sample thickness t.

The vortex melting scenario has been questioned by Moore\cite{ref26} who
argued that the melting transition could just be a second order crossover of
the vortex matter from a three dimensional behavior to a two dimensional one
when the l$_{c}$ in the vortex liquid becomes compatible to the sample
thickness {\it t}. Because {\it l}$_{c}$ grows very rapidly as the
temperature is lowered, the crossover region appears narrow enough to be
misinterpreted as a first order melting transition. For a cutting length of
the sample thickness {\it t}, the cutting temperature can be calculated
according to \cite{ref27}

\begin{equation}
B_{cut}=\frac{\Phi _{0}\varepsilon _{0}}{\gamma ^{2}Lk_{B}T}\text{,}
\label{eq1}
\end{equation}
by substituting L with {\it t}$\thickapprox $0.01 mm. Here $\phi _{0}$=2.07$%
\times $10$^{-7}$ Gauss cm$^{-2}$ is the flux quantum, $\varepsilon
_{0}=\Phi _{0}^{2}\ln \kappa /16\pi ^{2}\lambda _{ab}^{2}(T)$, $\kappa $
being the Ginzburg-Landau parameter and $\lambda _{ab}(0)\thickapprox 2000$%
\r{A}, the penetration depth in the ab plane. The obtained vortex cutting
line lies slightly below the melting line B$_{m}$(T). Although our result
can also be qualitatively explained with this argument, it is not clear why
such a sharp crossover is not observed in thin films (strong pinning
disorder) since in any sample with a finite thickness larger than the
distance between neighboring CuO planes, a crossover temperature is
expected. One reason could be that the relatively smaller numbers of pinning
centers in the single crystals (weak pinning) than in the thin films result
in a larger Larkin correlation length L$_{p}$ inside which an ordered VL can
persist.\cite{ref28} In the case of thin films, the transverse correlation
length is smaller than the average lattice spacing, only a second order
glass transition can be observed.\cite{ref29} This in turn suggests that a
regular VL with a relatively large radius is a necessity for the occurrence
of a melting transition. This is consistent with the fact that artificially
introduced pinning centers will induce a small Larkin length and smear out
the melting transition. Further work on the study of the relatiohship
between pinning strength and Larkin length need to be done in order to
clarify this point.

Our results suggest the following physical scenario: in a clean Y124 single
crystal, below the melting temperature T$_{m}$, the vortices form a rigid VL
with {\it l}$_{c}\gg t$. As the superconductor is warmed up, at T$_{m}$, VL
melting occurs. At this stage, the vortices lose their translational
coherence while keeping the correlations along the c-axis. Thus the vortices
in the temperature region just above the melting transition are in a 3D line
liquid state. At this temperature region, the vortex form a kind of
entangled vortices with their entanglement length {\it l}$_{c}$ smaller than
the sample thickness {\it t} but much larger than the distance s between the
CuO planes. The dissipation is governed by the vortex cutting and
recombination. As the temperature increases, the correlation length in the
c-axis becomes smaller and smaller and the difference between R$_{top}$ and R%
$_{bot}$ gets larger and larger. Finally at the temperature T$_{d}$,
inter-plane decoupling occurs ({\it l}$_{c}$ 
%TCIMACRO{\TEXTsymbol{<} }%
%BeginExpansion
\mbox{$<$}%
%EndExpansion
{\it s}) and the vortices lose the coherence in the c direction and the
vortices dissolve into 2D pancake vortices that move independently in the
individual CuO planes.

\section{Conclusions}

In conclusion, we have observed the melting transition in Y124 single
crystals. The vortices following the melting transition are in the 3D line
liquid state with a correlation length in the c- direction {\it l}$_{c}$
smaller than the sample thickness but larger than the distance between the
CuO planes of Y124. The temperature at which a resistive peak exists at R$%
_{bot}$ is found to correspond to the interlayer decoupling transition
temperature. Above the decoupling temperature, the vortices lose their
coherence in the c-direction and the dissipation would be governed by
tunneling of 2D vortices across the CuO planes. Our results also suggest
that the interlayer decoupling transition is a continuous crossover rather
than a sharp transition.

\section{Acknowledgments}

We thank E. Rossel and P. Wagner for their help during the measurements and
Y. Bruyneseraede for helpful discussions. This research has been supported
by the ESF Programme ''VORTEX'', the Belgian IUAP and Flemish GOA and FWO
Programmes.

\end{document}